\documentstyle[twocolumn,aps,prl,graphics,epsfig]{revtex}
\begin{document}
\title{Grover's search algorithm: An optical approach}
\author{P. G. Kwiat, J. R. Mitchell, P. D. D. Schwindt, and A. G. White}
\address{Physics Division, P-23, Los Alamos National Laboratory,
Los Alamos, New Mexico 87545}

\date{
To appear in a special issue of the Journal of Modern Optics--``The 
Physics of Quantum Information"}
\maketitle
\begin{abstract}
The essential operations of a quantum computer can be 
accomplished using solely optical elements, with different 
polarization or spatial modes representing the individual qubits.  We 
present a simple all-optical implementation of Grover's algorithm for 
efficient searching, in which a database of four elements is searched 
with a single query.  By `compiling' the actual setup, we have reduced 
the required number of optical elements from 24 to only 12.  We 
discuss the extension to larger databases, and the limitations of 
these techniques.\\
\end{abstract}

\noindent
{\bf 1. Introduction}\\
\\
It is now well-known that quantum computation, if implementable, could 
enable tremendous improvements over traditional computing methods for 
certain types of problems, such as factoring large numbers into primes 
\cite{Shor}, and efficiently searching a database \cite{Grover}.  Thus 
far, the implementation of actual algorithms has been limited to 
schemes employing bulk NMR methods \cite{NMR,Chuang}, although basic 
gate operations have also been performed using cooled ions \cite{ions} 
and also cavity QED methods \cite{QED}.  To date there have been a 
number of discussions on the implementation of basic quantum 
operations using purely optical methods 
\cite{Reck,Takeuchi,Summhammer,Cerf,Spreeuw}.  The central idea is 
that individual qubits can be represented by different polarization or 
spatial-mode degrees of freedom.  The difference from a genuine 
quantum computer with {\it distinct} entangleable registers is that 
the optical implementation requires a number of elements which grows 
exponentially with the number of qubits.  Nevertheless, as we shall 
see, it is possible to `compile' the circuits, allowing one to easily 
realize and test nontrivial algorithms involving several bits.  One 
conclusion of this approach is that a quantum computer is essentially 
a (complicated) interferometer \cite{Ekert}.

As shown in \cite{Cerf}, with this representation all the basic 
circuit elements of quantum computation can be accomplished using only 
linear passive optical elements.  This includes the Controlled Not 
(CNOT) gate, which entangles different bits, and the simpler 
Walsh-Hadamard (WH) transform: 0$\rightarrow$ (0+1)/$\sqrt{2}$ and 
1$\rightarrow$ (0$-$1)/$\sqrt{2}$.  For the present work we shall only 
need the latter, in addition to the ability to apply various phase 
shifts on the bits.  If one considers a qubit based on the 
polarization degree-of-freedom, where horizontal (H) and vertical (V) 
polarization represent 0 and 1, respectively, then the WH 
transformation can be accomplished with a half waveplate (HWP) 
oriented at $22.5^{\circ}$ to the horizontal.  Similarly, a simple 
50-50 beamsplitter (BS) performs the WH transform on a 
right-propagating spatial mode ($\equiv 0$) and an upward-propagating 
spatial mode ($\equiv 1$), after two extra phase shifters are 
included~\footnote{The BS transformations are well-known: $a 
\rightarrow (a+ib)/\sqrt{2}$; $b \rightarrow (ia+b)/\sqrt{2}$.  A 
$-\pi/2$ phase shift in the $b$ mode before and after the BS yields 
the WH transform\cite{Cerf}.}.  We will now describe how a nontrivial 
quantum circuit -- Grover's algorithm for efficiently searching a 
database -- may be constructed using these elements.\\
\\
\noindent
{\bf 2. Grover's Search Algorithm}\\
\\
\noindent
{\bf 2.1. General Description}\\
Consider a database of $N$ elements (e.g., a list of integers) exactly 
one of which is `marked' as satisfying some desirable characteristic 
(e.g., is a prime number).  Typically, one would expect to have to 
look at and test roughly half the database (making $\sim N/2$ 
`queries' on average) before locating this element.  Grover's 
algorithm uses the parallelism afforded by quantum superposition to 
accomplish the task with only $\sim \sqrt{N}$ queries.  Since it is 
described in detail elsewhere \cite{Grover,Chuang,Fahri}, here we 
present without proof the necessary steps.  Enough qubits are input to 
the computer to encode the elements of the database --- $n$ qubits 
suffice for a $2^n$-element database.  To initialize the computer, a 
WH transformation is performed on each qubit individually, thereby 
preparing the overall state into an equal superposition of each of the 
database elements, e.g., ($|00\rangle + |01\rangle+ |10\rangle + 
|11\rangle$)/2.  Next, an `Oracle" simultaneously examines all 
database elements, and marks one (or more) of them with a $\pi$ 
phase-shift.  In general, the Oracle might perform some computation on 
each element (e.g., test whether it is prime), and mark only that 
element which yielded a certain result (e.g., `prime").  For our 
proof-of-principle demonstration, the Oracle merely marks one element 
of the database (e.g., the state $|01\rangle$ if the second element is 
the desired one), while leaving the others unchanged.  Viewed as a 
computation, our Oracle accepts a user-specified input (e.g., `2'), 
and marks which database element matches that number.

Finally, a series of transformations accomplish the `inversion about 
the mean' operation at the heart of the amplitude-enhancement 
technique introduced by Grover \cite{Grover}.  These are:  (1) apply a 
second WH transformation to each bit; (2) apply a $\pi$ phase-shift to 
all but the first element ($|00\rangle$) of the database; and (3) 
again apply a final WH transform to each bit.  The net result of these 
last three operations is to transfer some of the amplitude from the 
non-special elements to the marked one.  Remarkably, for $N=4$ (i.e., a 
2-bit database), {\it all} of the amplitude is transferred to the 
desired element in a {\it single} run of the quantum circuit.  In 
general for larger $N$, running the circuit $\sim \pi \sqrt{N}/8$ 
times will lead to a final state in which the magnitude of the marked 
element's amplitude is greater than $1/\sqrt{2}$ -- a measurement on 
the system will then yield this element more than half the time 
\cite{Boyer}.  For our experiment we focused on the simplest case of a 
2-bit, 4-element database.  Below we discuss the extension to more 
bits.\\
\\
\begin{figure}
\begin{center}
\epsfxsize=\columnwidth
\epsfbox{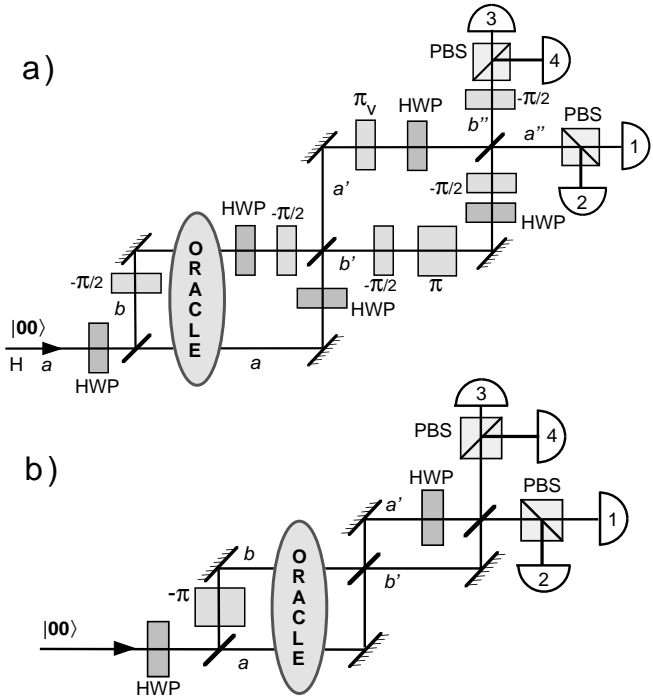}
\end{center}
\caption{Grover's technique for efficiently searching a database.
(a) A one-to-one optical implementation of the circuit,
for the case of n=2 bits (one represented by
polarization and the other by spatial mode.) All half waveplates (HWP)
are oriented at $22.5^{\circ}$. (b)  A `compiled' version of the
algorithm, where many of the components have been consolidated. The
first and second HWPs are at $22.5^{\circ}$ and $45^{\circ}$,
respectively.}
\label{Fig-1}
\end{figure}
{\bf 2.2. Optical Coding}\\
Figure 1a shows the optical layout corresponding to a direct 1-to-1 
implementation of Grover's algorithm, in which each operation is 
implemented independently.  A HWP oriented at $22.5^{\circ}$ performs 
the initial WH transformation on the polarization bit, while the WH 
transformation on the spatial-mode bit is performed by the first 50-50 
BS and the $-\pi/2$ phase shifter in the reflected path.  The oval 
represents the Oracle.  Below we will discuss a practical 
implementation of this device; for the moment consider simply 
inserting a waveplate in the spatial path $a$ or $b$, which depending 
on its orientation gives a $\pi$ phase shift to either H or 
V-polarization -- the Oracle can thus flip the sign of any one of the 
database elements $|aH\rangle$, $|aV\rangle$, $|bH\rangle$, or 
$|bV\rangle$.  For example, if the second element is the marked one, 
the state of the system after the Oracle is given by $(|aH\rangle - 
|aV\rangle+ |bH\rangle + |bV\rangle)/2 \equiv$ $(|00\rangle - 
|01\rangle + |10\rangle + |11\rangle)/2$.  Note that this is an {\it 
entangled} state of the two qubits -- it cannot be factorized into a 
product of a term involving only the polarization and a term involving 
only the spatial path.

The first WH transformation of the `inversion-about-the-mean' 
subroutine on the polarization bit is again accomplished by HWP's, and 
on the spatial mode by combining paths $a$ and $b$ on a second 50-50 
BS (with associated -$\pi/2$ phase shifters).  The minus sign on all 
but the $|00\rangle (\equiv |a'H\rangle$) element is obtained by a 
$\pi$-phase shift thickness of non-birefringent glass in path $b'$ (or 
by simply lengthening path $b'$ by $\lambda/2$), and a half waveplate 
in path $a'$, oriented with its fast axis horizontal (and labelled 
`$\pi_{V}$' in figure 1a).  The final WH transformations are again 
produced with HWP's and a 50-50 BS. We examine each of the outputs 
with a polarizing beamsplitter (PBS) in the H-V basis.\\
\\
{\bf 2.3. `Compiling'}\\
This basic optical circuit for Grover's algorithm may be simplified 
considerably by consolidating some of the optical transformations.  
For example, because we are free to choose the order in which the WH 
transforms are applied to the polarization and spatial modes (i.e., 
these operations on different qubits commute), we may move the HWP's 
in paths $a$ and $b$ after the second BS, and combine them with 
elements in the second interferometer --- the HWP-$\pi_{V}$-HWP 
combination in path $a'$ is equivalent to a single HWP oriented along 
$45^{\circ}$, and the HWP-HWP combination in path $b'$ yields the 
Identity transformation, i.e., no element at all.  After a few other 
consolidations of various phase shifts, the optical circuit of figure 
1a may be simplified to that in figure 1b.

To understand the device performance, we calculate, for example, the 
probability that a photon will exit to detector 1, by summing the 
amplitudes of the four possible paths, and taking the absolute square:
\begin{eqnarray}
\rm{P}(1)&=&|
\frac{1}{\sqrt{2}}\frac{1}{\sqrt{2}}A_{aV}
\frac{1}{\sqrt{2}}\frac{1}{\sqrt{2}}+
\frac{1}{\sqrt{2}}\frac{i}{\sqrt{2}}A_{bV}(-1)
\frac{i}{\sqrt{2}}\frac{1}{\sqrt{2}}\nonumber  \\
&+&\frac{i}{\sqrt{2}}\frac{i}{\sqrt{2}}A_{aH}
\frac{1}{\sqrt{2}}\frac{1}{\sqrt{2}}+
\frac{i}{\sqrt{2}}\frac{1}{\sqrt{2}}A_{bH}(-1)
\frac{i}{\sqrt{2}}\frac{1}{\sqrt{2}}
|^2\nonumber  \\
&=&\frac{1}{4}\left|
A_{aV}+A_{bV}-A_{aH}+ A_{bH}
\right|^2\;,
\end{eqnarray}
where the $A$'s represent the action of the Oracle, and each path's 
amplitude is calculated starting at the {\it output} (i.e., the first 
factor in each term represents transmission or reflection at the last 
non-polarizing beamsplitter).  The Oracle adds a relative phase of 
$\pi$ to the amplitude associated with the marked element of the 
database, while leaving the others unchanged.  We see that for the 
condition $\{A_{aH}$=-1; $A_{aV}$=$A_{bH}$=$A_{bV}$=1$\}$, P(1)=1; 
otherwise, if any of the other amplitudes received the $\pi$ phase 
shift instead, P(1)=0.  Thus, a click at detector 1 unambiguously 
determines that the marked database element was $|00\rangle$.  
Similarly, one can show that P(2)=1 for $A_{aV}$=-1, P(3)=1 for 
$A_{bH}$=-1, and P(4)=1 for $A_{bV}$=-1.\\
\\
\begin{figure}
\begin{center}
\epsfxsize=\columnwidth
\epsfbox{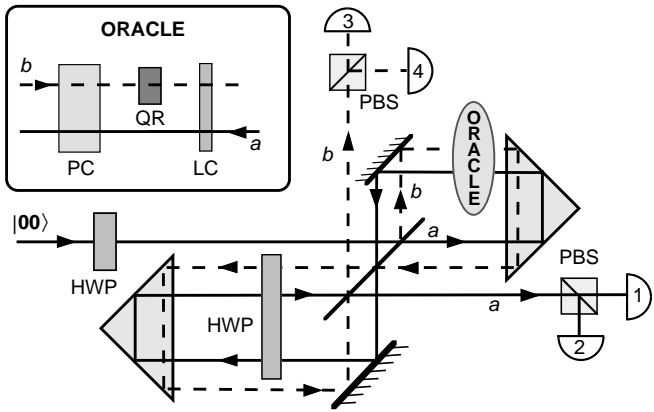}
\end{center}
\caption{Schematic of the self-stabilized optical setup used to
implement Grover's algorithm. Inset: Oracle.}
\label{setup}
\end{figure}
{\bf 3. Experimental Realization}\\
\\
{\bf 3.1. Stabilization and the `Oracle'}\\
One experimental constraint in any practical implementation is that 
the interferometer path lengths must be kept equal modulo $2\pi$ to 
have the desired transformation.  Instead of incorporating active 
stabilization, we chose to use a more robust interferometer geometry, 
based on displaced Sagnac interferometers.  In the usual Sagnac 
interferometer, the interfering paths arise from light propagating 
clockwise (CW) and counter-clockwise (CCW) through the system; because 
the optical path lengths are identical for both directions, the 
interferometer is automatically stabilized.  The disadvantage is that 
one cannot independently address the CW and CCW modes -- any optical 
element in one mode is automatically in the second as well.  Using a 
displaced geometry we can maintain the stability advantages of the 
usual Sagnac design, along with the ability to independently affect 
the CW and CCW modes.  The resulting experimental setup is shown in 
figure 2~\footnote{In actuality, the HWP (oriented at $45^{\circ}$) 
in the second interferometer acts on {\it three} of the paths instead 
of just one, ensuring that a photon horizontally-polarized in one of 
the retro-reflectors will be vertically-polarized in the other.  Any 
birefringent phase-shift from the total internal reflections is thus 
the same for all four trajectories through the system, and factors out 
of the final result.  This is a practical example of ``bang-bang''
quantum control\cite{Viola}. Also, we used the beamsplitter and mirrors at 
closer to normal incidence to reduce polarization-dependences.}.

The inset to figure 2 shows one implementation of the Oracle, which 
requires no moving parts~\footnote{Oracles that rely on insertion 
or rotation of a waveplate would necessitate constant realignment of 
the interferometer, due to slight wedges and wavefront anisotropies 
typical in these elements.}, 
and which was used to obtain the data presented here.  It consists of 
a Pockels cell (PC) [Lasermetrics \#Q1059], which acts as a piece of 
glass when no voltage is applied, and as a HWP oriented with the fast 
axis horizontal (i.e., H$\rightarrow$H and V$\rightarrow$-V) when a 
voltage of ~3.9kV is applied; a liquid crystal phase retarder (LC) 
[Meadowlark \#LRC-200-700], which acts as a piece of glass when a 
voltage of 5.6V is applied, and as a HWP with horizontal fast axis 
when a voltage of 2.2V is applied; and a quartz rotator, which rotates 
any linear polarization by $90^{\circ}$.  Consider, for example, the 
transformations when the applied voltages are 0kV and 2.2V:
\begin{eqnarray}
aH\rightarrow aH\rightarrow aH\;;\;\;
bH\rightarrow bH\rightarrow bV\rightarrow -bV\nonumber  \\
aV\rightarrow -aV\rightarrow -aV\;;\;\;
bV\rightarrow bV\rightarrow -bH\rightarrow -bH\;.
\end{eqnarray}
The input state $aH\equiv|00\rangle$' is the only one to acquire a net 
relative $\pi$ phase shift.  Similarly, the other three combinations 
of voltages on the Oracle components each apply the phase shift to a 
different element of the database.\\
\\
\begin{figure}
\begin{center}
\epsfxsize=\columnwidth
\epsfbox{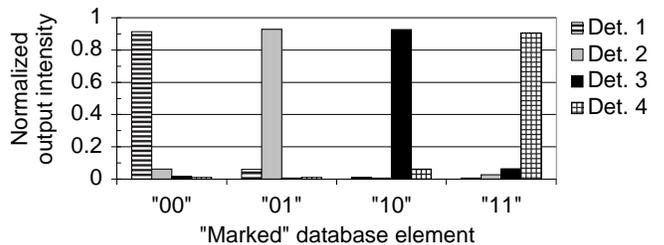}
\end{center}
\caption{Results obtained with the setup in figure 2.}
\label{data}
\end{figure}
{\bf 3.2. Results and Extensions}\\
The system in figure 2 was tested by directing a 
horizontally-polarized beam from a 670-nm laser diode into the 
interferometer, and monitoring the output intensities for the four 
different settings of the Oracle.  The results are shown in figure 3.  
The circuit essentially performed as expected --- the setting of the 
Oracle (i.e., the special element of the database) could be determined 
accurately with a {\it single} query~\footnote{While our experiment 
used many photons simultaneously for convenience, the identical 
results are predicted if single photons were used instead, i.e., only a 
single photon is needed to search the database.  Our laser beam may be 
thought of as many noninteracting copies of identical quantum 
computers, just as each individual molecule in the bulk NMR schemes is 
like a separate `computer', effectively isolated from the rest of the 
sample \cite{NMR}.}.  The average probability of error for a given 
port, as given by the relative output intensities, was less than 
2.8\%, and is mostly due to slight wavefront distortions introduced by 
the Oracle elements.

\begin{figure}
\begin{center}
\epsfxsize=\columnwidth
\epsfbox{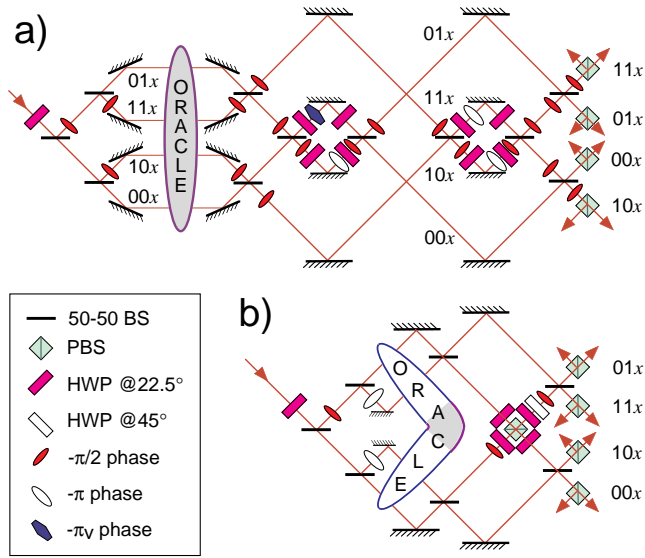}
\end{center}
\caption{Optical realization of Grover's algorithm for
n=3 bits.  A horizontally-polarized photon (polarization qubit
$\equiv$ `0').  The numbers represent the qubits of the spatial
modes at various stages of the computation; the {\it x} represents
the value of the polarization qubit. 
(a) The uncompiled version. (b) A compiled implementation 
(but not yet consolidated to make multiple use of redundant components).
A photon has a 25/32 = 0.78 probability to exit via the port 
corresponding to the database element marked by the Oracle, and a 1/32
probability to exit via each of the other seven outputs.}
\end{figure}
It should be possible to extend the current work to a greater number 
of bits $n$ and a correspondingly larger database.  One could use 
different frequencies of light, different orbital angular momentum 
modes\cite{Padgett}, or even different photon occupation numbers.  
However, the most 
straightforward extension is simply to add extra spatial modes (via 
beamsplitters), so that, for example, whether the photon is 
transmitted at the first BS defines the first bit, whether it is 
transmitted at the next BS becomes the second bit, and so on.  In 
figure 4a we show the direct optical coding of the Grover algorithm 
for a 3-qubit, 8-element database, requiring 63 optical elements.  One 
immediately sees the major limitation of these all-optical schemes, 
namely, that the number of optical elements grows as $\sim 2^n$ 
\footnote{Curiously, implementations using bulk NMR have a similar 
limitation, because the signal to noise for $n$ spins (bits) drops as 
$1/2^n$, due to the decreasing likelihood that all bits (which are in 
a thermal distribution) start off in the ground state \cite{NMR}.  In 
contrast, our photons start off in a very pure state.  Proposed 
schemes for ultracold NMR computers might also avoid this limitation 
\cite{Kane}.}.  However, this may be mitigated to some extent by 
`compiling' the algorithm (see figure 4b).  We have shown that, in 
fact, the requisite number of elements may be reduced to 21 or less, 
by making redundant uses of some of the components.
It is an open question how to achieve the maximal reduction, and to 
what extent reduction in an optical circuit implementation
corresponds to reduction with other `hardware'.\\
\\
{\bf 4. Discussion}\\
It is natural to ask what features of quantum computation are embodied 
in our all-optical implementation of Grover's algorithm.  In 
particular, we have seen that we can use different degrees of freedom
to represent individual qubits, and that we can readily prepare 
superposition states.  And although there are no CNOT gates in this
particular algorithm, the conditional phase shifts applied in the
amplitude-enhancement procedure, and by the Oracle itself, serve to
entangle the different degrees of freedom.  That one can so easily
achieve such entanglement is one of the 
advantages of the all-optical implementations of quantum circuitry.  
It is related to the fact that the optical implementation is much 
easier for operations that transform a given state of the database 
(e.g., applying a conditional phase shift to one element), because each 
state corresponds to a different physical mode of the system.  
Conversely, operations that transform a given {\it bit} become very 
difficult, requiring $\sim 2^{n-1}$ optical components (e.g., to flip 
the polarization qubit requires a HWP in half of the paths).  Contrast 
this with the standard multi-particle implementations of quantum 
computers, in which operations at the bit-level are more natural, 
while those on the overall states are among the most difficult.

There has been some confusion and controversy as to what extent a
state such as that generated by the Oracle is truly entangled.  On the
one hand, unlike more traditional multi-particle entangled states, an
entanglement in multiple degrees of freedom could not be used to 
demonstrate nonlocality, as in tests of Bell's inequalities.
The reason, however, is {\it not} that the states could not violate
a suitable Bell's inequality -- they can~\footnote{Because the 
polarization and spatial mode are commuting observables, they can both be 
determined simultaneously, the former with a simple polarizer,
the latter with a (non-polarizing) beamsplitter of arbitrary 
reflectivity.}; rather, it is that there is no way to achieve the 
space-like separation of the entangled systems necessary to satisfy the
underlying assumptions of Bell's inequalities.  So, in this sense our
entanglement is fundamentally different than multi-particle 
entanglement\cite{Spreeuw}.  On the other hand, the multiple 
degree-of-freedom states {\it do} satisfy the usual theoretical
criteria for entanglement\cite{Bennett}.  
For example, if we trace over one of the
systems (for instance, the polarization), the other system is left in a 
completely mixed state. Experimentally, this means, e.g.,, that 
the final polarization analysis is required for the interference to
be observed.

An interesting possibility for extending the capabilities of the 
present techniques is to incorporate the true entanglement \cite{BBO} 
(and even {\it hyper}-entanglement in several degrees of freedom 
\cite{hyper}) that exists between correlated photons produced via 
spontaneous parametric down-conversion (or, indeed, via other quantum 
computer embodiments).  For instance, it was recently shown that one 
could distinguish all four polarization Bell states using only passive 
linear elements, if the photons were simultaneously entangled in 
energy or momentum~\cite{embedded}.

Another major advantage of the all-optical methods is the virtual lack 
of decoherence, stemming from the fact that there are no preferred 
bases (e.g., the H/V polarization basis is not preferred to 45/-45).  
However, we can produce an {\it adjustable} decoherence simply by 
increasing the interferometer path-length imbalance relative to the 
coherence length of the light.  This technique was recently used to 
prepare an arbitrary mixed state of polarization \cite{Duality}, and 
to investigate optically various schemes for quantum control of 
decoherence \cite{Viola,qcontrol}.

Finally, we note that by combining these all-optical techniques with 
the ideas of `interaction-free measurements'~\cite{IFM,KwiatIFM}, one 
can demonstrate a remarkable prediction by Jozsa that a quantum 
computer can yield an answer without ever actually running 
\cite{Jozsa}!  The simplest version is to insert our entire circuit 
(minus the PBS's and detectors) into one arm of a Mach-Zehnder 
interferometer, such that the mode from the empty arm of the 
Mach-Zehnder is recombined with the `00/01' spatial-mode output of 
our system.  A {\it single} horizontally-polarized photon is input to 
the Mach-Zehnder, and the length of the empty arm is adjusted to give 
complete destructive interference for one output port of the 
recombining BS, when `00' is the marked element of the database (see 
figure 5a).  

If instead `01' is the marked element, 
a photon exiting our Grover circuit would be 
vertically-polarized, so no interference occurs at the recombining BS: 
a detector in the previously `dark' output port now fires half the 
time.  If the photon's polarization is vertical (1/2 of the time), we 
know the photon took the path containing the quantum computer (i.e., 
the Grover circuit `ran') and that `01' was the marked element.  
However, 1/4 of the time, the photon exiting the `dark' port will be 
horizontally-polarized (figure 5b).  
In this case, we know the marked element is 
{\it not} `00' (otherwise the destructive interference would have 
kept this port `dark'), and that {\it the photon absolutely did not 
take the interferometer path containing the Grover circuitry}.
If it had, either the photon would have vertical polarization
(if the special element were `01') or the photon would have left the
Grover system by its other output port (if the special element
were `10' or `11').  
Hence, 1/4 of the time we will have answered the question ``Is the 
marked element of the database not `00'?''  without ever running the 
computer.  By incorporating high-efficiency schemes for 
interaction-free measurements \cite{KwiatIFM}, one can in principle 
arbitrarily decrease the probability that the Grover circuitry is 
actually run if the marked element is not `00'. Note this is
an intrinsically quantum mechanical effect, relying on the 
indivisibility of a single photon.  Note also that although the photon
may not have actually traversed the search algorithm optics, their
alignment is still critical, and, for example, the electrooptic 
Oracle elements still require power.
\begin{figure}
\begin{center}
\epsfxsize=\columnwidth
\epsfbox{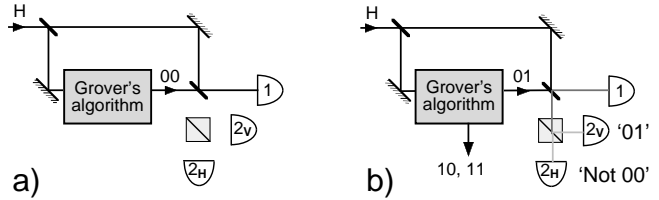}
\end{center}
\caption{Incorporating `interaction-free' measurements.  (a) The 
interferometer is adjusted so that port 2 has complete destructive
interference if the Oracle setting is `00'. (b) If the Oracle 
setting is `01', then the exiting photon has vertical polarization,
and no interference occurs -- a detector in port 2 can now fire,
indicating that the Oracle setting is not `00'; 
half the time the photon in port 2 will nevertheless still have horizontal 
polarization, indicating that the quantum algorithm did not actually
run.}
\end{figure}
In summary, we have demonstrated an all-optical realization of 
Grover's search algorithm.  Because the qubits in our system are 
represented by different degrees of freedom, instead of separate 
quanta, many of the gate operations are much simpler to implement.  We 
stress that our results are completely different than simply 
performing a (digital) calculation to predict the behavior -- ours is 
a {\it physical} system that relies on superposition, interference, 
and non-factorizable states to function.  Since these may also be 
classical phenomena, we conclude that many ingredients of quantum 
algorithms are not necessarily non-classical.  The role of `true' 
entanglement is to provide an exponential savings in resources.  
However, by compiling the optical circuity, algorithms involving 
several bits may be readily investigated.  \\
\\
{\bf Acknowledgements}
\\
We would like to acknowledge P. Wang and E. Waks for their assistance, 
and R. Jozsa for helpful discussions.

\end{document}